# THE EVOLUTION OF ACCRETION DISKS WITH CORONAE: A MODEL FOR THE LOW-FREQUENCY QUASI-PERIODIC OSCILLATIONS IN X-RAY BINARIES

Marek A. Abramowicz and Xingming Chen

Department of Astronomy and Astrophysics, Göteborg University and Chalmers University of Technology, 412 96 Göteborg, Sweden

marek@fy.chalmers.se; chen@fy.chalmers.se

and

Ronald E. Taam

Department of Physics and Astronomy, Northwestern University, Evanston, IL 60208

taam@ossenu.astro.nwu.edu

## ABSTRACT

The global nonlinear time dependent evolution of accretion disk-corona systems in X-ray binary sources has been investigated to provide an understanding of the low frequency ($\sim 0.04$ Hz) quasi-periodic oscillations (QPOs) observed recently in the Rapid Burster MXB 1730-335 and in some black hole candidates sources (Cyg X–1 and GRO J0422+32). We consider $\alpha$-viscosity models in which the viscous stress is proportional to the total pressure. In contrast to previous time dependent studies it is assumed that all mass accretion and angular momentum transport take place in an optically thick disk, but that a fraction of the gravitational energy that is released is dissipated in a corona. It is found that the coronal energy dissipation can effectively reduce the amplitudes of the mass flow variations generated from the thermal and viscous instabilities (in comparison to models without a corona). Provided that the disk is close to a marginally stable state mild oscillatory nonsteady behavior results. These oscillations are globally coherent in the unstable regions of the disk.

A model for the high and low states of black hole candidate systems is also proposed. It is suggested that the low state, which is characterized by a hard X-ray spectrum, corresponds to a disk configuration in which the inner disk is in an advection dominated hot optically thin state whereas the high state corresponds to a configuration in which the inner disk is in an optically thick state surrounded by a corona. In this model, the mass accretion rate in the system is higher in the low state than in the high state. The hard X-ray spectrum of QPOs observed in the low state can be naturally explained by such a model.

*Subject headings:* accretion, accretion disks — black hole physics — instabilities

## 1. INTRODUCTION

The spectral behavior of galactic black hole candidates (BHCs) and active galactic nuclei (AGNs) contains valuable information about the underlying physics of the accretion process. It is now well accepted that the soft X-ray black body spectral component from BHCs and the UV black body radiation from AGNs is due to a cold optically thick accretion disk. On the other hand, the power-law spectra from these sources suggest that optically thin hot matter is also present. The quasi-periodic oscillation (QPO) phenomenon in low mass X-ray binaries (LMXBs) and BHCs also provides a probe of the accretion process in these systems. In particular, an understanding of the observed luminosity variations may place important constraints on the physics of accretion in these objects. The QPOs in LMXBs are classified as horizontal-branch and normal-branch QPOs with frequencies in the range of $\sim 20 - 50$ Hz and $5 - 10$ Hz respectively. The models proposed for these QPOs involve neutron stars (for a review see van der Klis 1989). Recently, low frequency QPOs with characteristic frequencies of $\sim 0.04$ Hz have been observed from the Rapid Burster MXB 1730–335 (Lubin et al 1992), Cyg X-1 (Angelini & White 1992; Vikhlinin et al 1992a, 1994; Kouveliotou et al 1992a), and GRO J0422+32 (Vikhlinin et al 1992b, Kouveliotou et al 1992b, Pietsch et al 1993). In the Rapid Burster, the QPOs are observed in the soft X-ray energy band, whereas in BHCs the QPOs are observed in the hard X-ray spectral state of these sources. In the hard state of BHCs the source spectrum is characterized by a single power law and strong flickering activity (see Tanaka 1991). The mechanism responsible for the production of these low frequency QPOs may be different from those proposed for the other classes of QPOs since these oscillations have been observed from both neutron star and black hole candidate systems. Milsom, Chen, & Taam (1994) and Chen & Taam (1994) recently proposed a disk instability model in which the thermal and viscous modes of the accretion disk are mildly unstable based upon a modified viscosity prescription, $\tau = -\alpha_0 (H/R)^n p_g$, where $\tau$ is the viscous stress, $p_g$ is the gas pressure, $H$ is the local scale height of the disk, $R$ is the distance from the central object, and $\alpha_0$ and $n$ are constants. This model was motivated by the fact that accretion disks based on a viscous stress proportional to the total pressure are unstable (Lightman & Eardley 1974) producing large-amplitude, burst-like fluctuations (e.g., Taam & Lin 1984; Lasota & Pelat 1991; Honma, Matsumoto, & Kato 1991), whereas accretion disks based on a viscous stress proportional to the gas pressure are always thermally and viscously stable.

More recently, Chen (1995b) suggested that a disk-corona system is naturally weakly unstable due to the stabilization effect associated with coronal energy dissipation confirming earlier work of Ionson & Kuperus (1984) and more recently by Svensson & Zdziarski (1994). The disk-corona configuration is one of the general scenarios proposed to account for the presence of optically thin hot matter in X-ray binary systems (see, e.g., Liang & Price 1977). The concept of a corona has also been invoked in the interpretation of the spectra from Seyfert galaxies (see, e.g., Haardt & Maraschi 1991). Consequently, theoretical studies of the structure of a disk-corona system have been undertaken by a number of groups including Nakamura & Osaki (1993), Kusunose & Mineshige (1994) and Svensson & Zdziarski (1994). The inclusion of a corona leads to a reduction in the ratio of radiation pressure to gas pressure within the optically thick disk as the fraction of the energy dissipated in the corona is increased. Hence, the mass accretion rate threshold above which the disk is unstable to thermal and viscous perturbations is raised (see Ionson & Kuperus 1984).

In this paper, we report on the results of the time-dependent evolution of accretion disk-corona systems and investigate their possible relevance to the low frequency QPOs observed in LMXBs. In the next section we outline the assumptions and approximations underlying the model calculations. The numerical results of the calculations are presented in §3. Finally, the hard spectral characteristics of the observed QPOs in BHCs are incorporated into the framework of our disk instability model to provide a model for the high and low states of black hole transient binary systems in the final section.

## 2. FORMULATION

We consider a disk-corona system which is axisymmetric and non self-gravitating. The disk is optically thick and geometrically thin so that it can be described by the vertically integrated equations. The overlying corona is hot and optically thin. Following Svensson & Zdziarski (1994) it is assumed that all the mass accretion and angular momentum transport take place in the cold optically thick disk. We neglect the radiation drag which may be important in transporting angular momentum in the corona as in models of normal branch oscillations (Fortner, Lamb, & Miller 1989) in the bright neutron star X-ray binaries. In the case of BHCs, the importance of such effects is less clear. Since we focus on the thermal-viscous instability and do not consider the inertial-acoustic instabilities which have a much shorter timescale and may exist possibly in the innermost regions of the disk (see, e.g., Matsumoto, Kato, & Honma 1989; Nowak & Wagoner 1992; Wallinder 1991; Chen & Taam 1993, 1995), a Keplerian disk is assumed. In these approximations, the surface density of the optically thick disk, $\Sigma$, at a given cylindrical radius, $r$, can be obtained by combining the conservation equations of mass and angular momentum. It is given by the standard time dependent mass diffusion equation (Lynden-Bell & Pringle 1974) as

$$\frac{\partial \Sigma}{\partial t} = \frac{3}{r} \frac{\partial}{\partial r} \left[ r^{1/2} \frac{\partial}{\partial r} \left( \nu \Sigma r^{1/2} \right) \right], \tag{1}$$

where $t$ is the time, and $\nu$ is the effective kinematic viscosity. In equation (1), the Newtonian Keplerian angular velocity, $\Omega_k = \sqrt{GM/r^3}$, is implicitly assumed, where $M$ is the mass of the central compact object. Relativistic effects on the gravitational potential are important near the inner edge of the disk, but we anticipate that the long timescale ($\sim 25$ s) associated with the low frequency QPOs dictates that the region of instability lies at sufficiently large radii compared to the innermost disk that the results we obtain will be insensitive to the neglect of these effects. In addition, the lack of a strong soft X-ray emission component during the phases when QPOs are observed in BHCs suggest that an optically thick disk is not present in the innermost regions. The viscosity is described by the standard $\alpha$-model prescription as (Shakura & Sunyaev 1973):

$$\nu = \frac{2}{3}\alpha c_s H, \tag{2}$$

where $\alpha$ is a constant, $c_s = \sqrt{P/\Sigma}$ is the local sound speed, and $H = c_s/\Omega_k$ is the half-thickness of the disk. Here $P$ is the vertically integrated total pressure of the disk.

The energy conservation equation for the cold accretion disk is represented by the balance between the local viscous heating, the local radiative cooling, and the global heat transport (radial advection). It is expressed as

$$P\frac{4-3\beta}{\Gamma_3-1}\left[\left(\frac{\partial \ln T}{\partial t} + v_r \frac{\partial \ln T}{\partial r}\right) - (\Gamma_3 - 1)\left(\frac{\partial \ln \Sigma}{\partial t} + v_r \frac{\partial \ln \Sigma}{\partial r} - \frac{\partial \ln H}{\partial t}\right)\right]$$
$$= F^+ - F^- - \frac{2}{r}\frac{\partial(rF_r H)}{\partial r}, \tag{3}$$

where $T$ is the mid-plane disk temperature, $\beta = P_g/P$ is the ratio of the gas pressure to the total pressure, $\Gamma_3 = 1 + (4-3\beta)(\gamma-1)/[\beta + 12(\gamma-1)(\beta-1)]$, and $\gamma = 5/3$ is the ratio of specific heats. Here, the radial velocity, $v_r$, is given as

$$v_r = -\frac{1}{\Sigma r^{1/2}}\frac{\partial(\nu \Sigma r^{1/2})}{\partial r} \tag{4}$$

and

$$\dot{M} = 2\pi r \Sigma v_r. \tag{5}$$

Although eq(3) is cast in a slightly different form than in previous studies, it is equivalent to the energy equation used in Chen & Taam (1994). Here, heating from the corona has not been included (see Svensson & Zdziarski 1994). This is justified by the fact that the heating of the disk by the corona only takes place in the surface layers down to a few Thompson optical depths (see Ross & Fabian 1993). Furthermore, it is known that the effect of irradiation is unimportant in the deep interior of the disk unless the irradiated flux is comparable to the product of the viscously generated flux and the optical depth of the disk (Lyutyi & Sunyaev 1976; Taam & Mészáros 1987). We assume that a fraction $\eta$ of the total gravitational potential energy released is dissipated in the corona (Haardt & Maraschi 1991). Since the physics of the dissipation in the corona is not well understood, $\eta$ is left as a free parameter. Thus, within the disk, the viscous heating rate per unit area becomes

$$F^+ = (1-\eta)\nu\Sigma\left(r\frac{\partial \Omega_k}{\partial r}\right)^2. \tag{6}$$

In the optically thick radiative diffusion approximation, the local radiative cooling rate in the vertical direction is

$$F^- = \frac{4acT^4}{3\kappa\Sigma}, \tag{7}$$

where $\kappa$ is the sum of electron scattering and free-free opacities. The radiative energy flux in the radial direction is written as

$$F_r = -2HF^-\frac{\partial \ln T}{\partial r}. \tag{8}$$

Finally, for optically thick accretion disks, the equation of state is given by

$$P = \frac{\mathcal{R}\Sigma T}{\mu} + 2H\frac{aT^4}{3}, \tag{8}$$

where $\mu$ is the mean molecular weight assumed to be 0.617.

The optically thick disk is assumed to be truncated at an inner radius $r_{in}$ to allow for the possible existence of an inner optically thin region. We adopt such a picture to be consistent with the spectral data during the QPO active phase (i.e., the presence of a hard X-ray spectrum and the lack of a significant strong soft X-ray component). Thus, we postulate that the oscillations which originate in an optically thick outer cold disk, propagate to an optically thin hot inner disk and produce the observed X-ray QPO variability.

The time dependent equations (1) and (3) are solved to calculate $\Sigma$ and $T$ via an explicit method. The accretion disk is divided into 41 grid points distributed equally on a logarithmic scale ranging from $r_{in}$ to an outer boundary, $r_{out} = 50r_{in}$. The initial structure of the disk is given by the thermal equilibrium solutions (see Chen 1995b). The outer boundary of the disk is chosen to correspond to a fixed temperature and surface density to ensure that the flow of matter into the computational domain occurs at a fixed rate. At the innermost radius of the disk we impose the condition that the gradients of the temperature and surface density are zero.

## 3. NUMERICAL RESULTS

The accretion disk model is characterized by the mass of the central object, $M$, the mass accretion rate, $\dot{M}$, the viscosity parameter, $\alpha$, the coronal dissipation fraction, $\eta$, and the inner radius of the cold disk. We adopt the mass as $1.4\,M_\odot$ for neutron stars and $10\,M_\odot$ for black holes. For convenience, the mass accretion rate is expressed in units of the critical limit defined as $\dot{M}_c = 4\pi GM/(\kappa_{es}c\epsilon)$, where $\kappa_{es} = 0.34$ is the electron scattering opacity and $\epsilon$ is the efficiency for the conversion of rest mass energy into radiation taken to be 1/6 for a neutron star and 1/16 for a black hole. The radii are measured in units of the Schwarzschild radius defined as $r_g = 2GM/c^2$.

To delineate the phase space in $\dot{M}$, $\eta$, and $r_{in}$ for which the accretion disk is marginally stable, we have carried out time dependent calculations for a range of these parameters. The results are illustrated in Figure 1 for a black hole of mass $10M_\odot$ and $\alpha$ parameter equal to unity. For greater fractions of the gravitational potential energy dissipated in the corona, the disk is stabilized after a few oscillations (less then 4-5), but at lower values of $\eta$ than that indicated by linear stability analysis (see Ionson & Kuperus 1984; Svensson & Zdziarski 1994; Chen 1995b), due to the affects of advective energy transport. We point out that oscillations can occur from an initial steady state model due to the fact that the initial disk structure is not an exact solution of the time dependent equations. In particular, to construct the initial disk the radial diffusion term in the energy equation (see eq. 3) is not included and the advection term is only approximately considered (see Chen 1995b). These terms act as initial perturbations to the disk which lead to the transient temporal development and eventually to either stabilization or oscillation. From Fig. 1 it can also be seen that the critical value $\eta_c$, above which the disk is stable, is a function of the mass accretion rate. In particular, $\eta_c$ increases with the mass accretion rate for a given inner radius of the disk since a greater energy dissipation in the corona is required to compensate for a larger unstable region in the disk at higher mass accretion rates (in the absence of the energy dissipation in the corona). Furthermore, it can be seen that $\eta_c$ asymptotically approaches a limiting value for $\dot{M} \sim \dot{M}_c$ with the limiting value approached at lower mass accretion rates for smaller inner disk radii. For example, $\eta_c$ approaches a value of 0.59 to 0.88 at $\dot{M}$ ranging from $0.7 - 0.3\dot{M}_c$ for inner disk radii ranging from $70r_g$ to $10r_g$. It is also found that $\eta_c$ is greater for disks of smaller radii at a given mass accretion rate. This follows from the fact that the unstable region is larger for smaller inner disk radii, and, hence, greater energy dissipation in the corona is required to place the disk in the stable gas pressure dominant region.

The time dependent calculations reveal that the characteristics of the global oscillations resulting from the development of the thermal-viscous instability are sensitive to parameters which deviate from those delineated by each curve, especially for those at large radii. For example, the oscillations are of greater amplitude for lower $\eta$ since a larger fraction of the disk is involved in the outburst. In addition, the timescales of the global oscillations are generally longer for disks characterized by larger inner disk radii and smaller fractions of energy dissipated in the corona. For oscillation timescales in the observed range of the low frequency QPOs ($\sim 20 - 30$ s), parameters must be chosen which lie extremely close to a given curve. This would not likely to be the case in general and parameters deviating from these curves are expected. We point out that it is possible that $\eta$ may be affected by the instability in the disk. For example, for greater instability the coronal dissipation may be increased, thereby increasing $\eta$ and reducing the tendency toward instability. This may force the disk to always lie near the marginally stable state. Although this is plausible, it is difficult to quantify and, instead, we consider the simplest case that $\eta$ is not a function of the strength of the instability. In this case, it is found that significant changes in the properties of the oscillations would occur at large radii. On the other hand, the sensitivity of the results to the parameters deviating from the curve is much weaker for small inner disk radii (10 $r_g$ - 30 $r_g$) where the dissipation of gravitational potential energy in the corona must be greater to produce characteristics similar to the observed QPOs from BHCs. In the following we describe the numerical results from studies at these smaller inner disk radii.

In the first sequence (see Table 1), which we have denoted as the standard model, we have followed the evolution of the thermal instability in a disk surrounding a black hole of $10M_\odot$ accreting at $\dot{M} = 0.70\dot{M}_c$. It is found that the disk undergoes globally coherent oscillations. This behavior is illustrated in Fig. 2 where the evolution of the local temperature and column density is shown for four different radii. For reference we also include the steady state S curves which are locally thermally and viscously stable for $\frac{dT}{d\Sigma} > 0$ and unstable for $\frac{dT}{d\Sigma} < 0$. It can be seen that the evolutionary paths take the form of loops with larger deviations from the steady state structure present at smaller disk radii. Because of the importance of advective energy transport during the time dependent evolution, the paths in these diagrams during the unstable stages do not follow trajectories at constant $\Sigma$ (see also, Chen & Taam 1994), but form loops centered near the lower turning point of the S curve. Note also that the average temperature is slightly less than the initial steady state one, especially at small radii. This is due to the additional cooling terms included in the time-dependent energy equation, namely, the radial radiative diffusion and the global advection. For this sequence the disk remains steady for radii $\gtrsim 174 r_g$. The timescale of the resulting mass flow variations is calculated to be 28 s with an amplitude of $\frac{\Delta \dot{M}}{\dot{M}} \sim 30\%$ (see Table 1 and Fig. 3a).

To determine the sensitivity of the results to the mass accretion rate, $\dot{M}$ was changed to 0.65 and 0.75 in sequences 2 and 3 respectively, with all other input parameters held fixed as in the standard sequence. That is, we choose parameters which place the model disk nearly parallel to the marginally stable curve characterized by an inner disk radius of $30 r_g$. The disk is very weakly nonsteady as shown in Figure 3b for sequence 2, but the amplitude and period of the oscillation for sequence 3 has increased slightly (Fig. 3c). Hence, the frequency of the oscillation increases whereas the relative mass flow rate fluctuation amplitude decreases as the mass accretion rate is decreased. However, this relation holds only with other disk parameters held unchanged and only over a limited range in mass accretion rates for which the disk is nonsteady. The latter reflects the fact that for higher mass accretion rates ($\gtrsim \dot{M}_c$ for these cases) the disk becomes stable due to the effect of radial advection (for example, see the upper stable branch in Fig. 2)

The nonstationary behavior of the accretion disk is most sensitive to the coronal dissipation parameter, $\eta$. For example, if we choose parameters away from the vicinity of the marginally stable curve by decreasing $\eta$ in the standard sequence to 0.7, the region where the disk is locally unstable increases and, thus, the disk exhibits larger amplitudes and longer period oscillations (sequence 4, Fig. 4a). On the other hand, if $\eta$ is increased above $\eta_c$, i.e. if $\eta$ lies to the right of the curve at a given radius as illustrated in Fig. 1, the disk is stable.

The global instability of the disk also depends on the location of the inner edge of the cold disk. For example, by decreasing $r_{in}$ from 30 $r_g$ to 20 $r_g$, with other disk parameters held fixed, the spatial extent of the unstable region of the disk increases and the mass flow modulations tend to have a larger amplitude and lower frequency (sequence 5, Fig. 4b) with respect to the standard sequence. By considering the influences of $\eta$ and $r_{in}$ together, similar oscillation amplitudes and timescales to the standard sequence can be obtained with, for example, $\eta = 0.72$ and $r_{in} = 35 r_g$ (sequence 6, Fig. 4c).

Both these properties of the oscillations are determined by the parameters of the disk-corona system. To be consistent with the frequency and amplitude of QPOs from BHCs, our calculations suggest that, $\alpha \approx 1$, $\eta \approx 0.75$, $r_{in} \approx 30 r_g$, and $\dot{M} \approx 0.7 \dot{M}_c$ for a $10 M_\odot$ black hole. If the black hole is less massive, then $\alpha$ shall decrease approximately proportionally. We should also point out that, the result depends on $\dot{M}$ less sensitively for large $\eta$ then for smaller $\eta$. Thus a larger $\eta > 0.5$ is preferred to avoid excessive fine tuning. A small value of $\alpha$ leads to lower disk temperatures and increases the thermal and viscous timescales. In this case a smaller value of $\eta$ is required to destablize the disk. Thus we estimate $\alpha$ cannot be less than $\sim 0.5$.

To determine the dependence of the oscillations on the mass of the compact object we decreased $M$ to 1.4 $M_\odot$ to model the evolution of an accretion disk surrounding a neutron star. In order to produce oscillations at frequencies $\sim 0.04$ Hz as observed in the Rapid Burster, the viscosity parameter, $\alpha$, must be chosen to be smaller than the black hole case since the absolute size of the inner disk is smaller. The thermal and viscous time scales (relative to the dynamical time scale) would be correspondingly increased. Since a smaller $\alpha$ results in a cooler disk making it more stable, smaller values of both $\eta$ and $r_{in}$ are required to promote global instabilities in the disk. Figure 5 shows three examples which are relevant to the case of Rapid Burster. In all three sequences (7, 8, and 9), the fraction of energy dissipated in the corona is assumed to be 0.5. It can be seen that variability with frequencies in the range of 39 mHz to 64 mHz can be produced provided that the inner edge of the optically thick disk lies close to the neutron star surface. Specifically, the mass flow rate at $r_{in} = 8r_g$ in sequence 7 varies on a timescale of about 16 s with an amplitude of 50% for a mass input rate of 0.3 $\dot{M}_c$ and $\alpha = 0.02$. Variations of the input parameters lead to similar trends as described for the black hole case.

## 4. DISCUSSION

It has been shown that low amplitude mass flow variations can be produced in an optically thick accretion disk in which the viscous stress is assumed to be proportional to the total pressure. An essential feature in the model is the recognition that if a coronal region exists above the cold disk the nonlinear development of the radiation pressure induced thermal-viscous instability can be moderated. We have found that the stabilization of the disk depends not only on the amount of energy dissipated in the corona, but also on the location of the inner radius of the optically thick structure. Stabilization associated with $\eta$ had been noted by Chen (1995b, see also Ionson & Kuperus 1984 and Svensson & Zdziarski 1994) whereas the stabilization associated with $r_{in}$ has been shown in the present work. This latter tendency toward stabilization can be understood as resulting from the reduction in the width of the unstable region. For a sufficiently narrow unstable region the disk is stabilized by advective energy transport (e.g., Chen & Taam 1994). These independent trends can be combined to illustrate the tendency toward stabilization. For example, for an inner boundary chosen closer to the central object, a greater fraction of gravitational potential energy must be dissipated in the corona to stabilize the disk. The numerical calculations demonstrate that the mass flow variations can be of small amplitude provided that the fraction of energy dissipated in the corona, $\eta$, is high. Specifically mass flow modulations on a timescale of 20 - 30 s can be produced for $\eta \gtrsim 0.75$ for an inner boundary of the optically thick disk of $\lesssim 30r_g$. For $\eta \lesssim 0.75$ the amplitude variations increase in size ($\gtrsim 30\%$). The disk exhibits nonsteady behavior for a range in the mass accretion rates for a given $\eta$ and $r_{in}$. For example, for $\eta = 0.75$ and $r_{in} = 30r_g$ the disk exhibits mass flow modulations for $\sim 0.65\dot{M}_c - 1\dot{M}_c$ for $\alpha$ equal to unity.

QPO behavior is expected for a modulation of the mass input rate into the inner disk region. The variation of the QPO frequency with respect to mass accretion rate is nonmonotonic since the inner radius of the optically thick disk is likely to vary as well. For higher mass accretion rates it is likely that the inner disk radius increases and these variations are expected to lead to opposing tendencies. That is, for a fixed inner boundary and a given fraction of gravitational potential energy dissipated in the corona, the QPO frequency decreases with increasing mass accretion rate. On the other hand, the QPO frequency tends to increase for larger inner disk radii at fixed mass accretion rate (provided that the inner radius is not increased to the extent that the disk becomes steady). This suggests that the variations of the QPO frequency depend less sensitively on the mass accretion rate than in the case where the inner edge of the optically thick disk is fixed in radius.

The quasi-periodic oscillations observed from black hole candidate transient systems are characterized by a hard spectrum (e.g. Vikhlinin et al. 1992; Kouvelioutou et al. 1992). This indicates that the optically thin regions in the disk participate in the phenomenon. Such a disk structure is expected for $\alpha \sim 1$ since the inner regions of such disks would be become optically thin for $\dot{M} \gtrsim 0.1\dot{M}_{\rm c}$. This disk configuration is similar to the two temperature inner disk proposed by Shapiro, Lightman, & Eardley (1976). In the present context, however, we assume that the hot inner disk is stabilized by the advection of energy and entropy into the black hole as conceived by Abramowicz et al (1988, 1995), Kato, Honma, & Matsumoto (1988) Chen & Taam (1993), Narayan & Popham (1993), Narayan & Yi (1994, 1995a,b) and Chen (1995a, b). In contrast to the Shapiro et al (1976) picture, we also require a corona in the outer cold optically thick disk as well to moderate the thermal-viscous instability. Thus, in our model, the inner disk responds to the mass flow modulations produced by the thermal-viscous instability in the outer region. As a result, the QPOs will be characterized by a hard spectrum. Our model is also consistent with the ROSAT observation of QPOs of J0422+32 in soft X-rays by Pietsch et al. (1993) provided that $r_{in} \lesssim 30 r_g$ where the soft X-ray component can be produced in the optically thick disk.

The low frequency QPOs observed in the soft X-ray energy band from the Rapid Burster MXB 1730-335 likely result from the modulations of the disk accretion onto the neutron star surface. The lack of hard X-rays suggests that either the electron temperature characterizing the optically thin gas is lower than in BHCs and/or that the optically thick disk extends closer to the central object. The fact that the electron temperatures in the optically thin regions of the disk are lower for neutron star binaries is expected because of the additional cooling provided by the soft X-ray flux emitted from the neutron star surface (Sunyaev et al. 1991). The interpretation with regard to the presence of an optically thick disk close to the neutron star is also consistent with the fact that lower viscosities, which enable the optically thick disk to lie closer to the neutron star surface, are required to explain the timescale of the QPOs for lower mass systems. For these viscosities, the presence of an optically thin region is attributed to the existence of a small neutron star magnetosphere which disrupts only the very innermost regions of an optically thick disk.

The transition from an optically thick to an optically thin disk surrounding a black hole at $r_{in}$ can, in principle, be determined from theoretical considerations. It is known that there exists a relation between the mass accretion rate and disk radius for which an advection dominated optically thin solution is the only possible solution (Chen et al 1995) for accretion rates above a certain limit. On the other hand, for mass accretion rates below the limit, three solutions are possible of which two are stable (optically thick and advection dominated optically thin) and one is thermally unstable (optically thin solution in which local cooling is dominant). At these lower mass accretion rates, we suggest that the optically thick region extends to the black hole. Furthermore, we propose that the innermost regions of the disk are characterized by an optically thick structure surrounded by an optically thin hot corona. In this case, the hard X-rays are produced by a combination of the synchrotron process involving the relativistic electrons in the corona and the magnetic field in the disk and by inverse Compton scattering of the soft photons emitted from the cool disk. In this state, the soft X-ray photons result from the reprocessing of the hard X-rays produced in the corona which impinge on the cold disk and from the energy dissipated in the cold disk itself. We identify this state as the high state of BHCs. This interpretation is consistent with the observations of BHCs in which the soft X-ray spectrum is well fit by multicolor black body temperatures in the disk with an inner radius of 3 $r_g$ (Tanaka 1991; Tanaka & Lewin 1995) and with the fact that the X-ray spectrum exhibits short timescale variability primarily at higher energies (e.g., Ebisawa 1991; Ebisawa et al 1994). In this model, the low hard spectral state of BHCs, wherein a significant fraction ($\gtrsim 99\%$) of the energy generated in the optically thin disk is advected into the black hole (Chen 1995a; Narayan & Yi 1995b), occurs at higher mass accretion rates than the high soft spectral state. Hence, the bolometric luminosity characterizing the low state can be lower than the high state even though the mass accretion rate is higher.

Future theoretical investigations will be directed toward a more detailed study of the relation between the mass accretion rate and the inner radius of the optically thick disk. Such investigations are necessary for determining the relation of the QPO frequency with respect to the mass accretion rate and for delineating the mass accretion rate regime separating the high and low states of the BHCs.

This research was supported in part by NASA under grant NAGW-2526. RET thanks Dr. F. Haardt for discussions while he was a visitor in Göteborg.

# REFERENCES

Abramowicz, M. A., Chen, X., Kato, S., Lasota, J. P., & Regev, O. 1995, ApJ, 438, L37

Abramowicz, M. A., Czerny, B., Lasota, J. P., & Szuszkiewicz, E. 1988, ApJ, 332, 646

Angelini, L., & White, N. 1992, IAU Circ. No. 5580

Chen, X. 1995a, MNRAS, in press

Chen, X. 1995b, ApJ, in press

Chen, X., Abramowicz, M. A., Lasota, J. P., Narayan, R. & Yi, I. 1995, ApJ, in press

Chen, X., & Taam, R. E. 1993, ApJ, 412, 254

———. 1994, ApJ, 431, 732

———. 1995, ApJ, 441, 354

Ebisawa, K. 1991, PhD Thesis, Univ. of Tokyo

Ebisawa, K. et al. 1994, PASJ, 46, 375

Fortner, B., Lamb, F. K., & Miller, G. S. 1989, Nature, 342, 775

Haardt, F., & Maraschi, L. 1991, ApJ, 380, L51

Honma, F., Matsumoto, R., & Kato, S. 1991, PASJ, 43, 147

Ionson, J. A., & Kuperus, M. 1984, ApJ, 284, 389

Kato, S., Honma, F., & Matsumoto, R. 1988, MNRAS, 231, 37

Kouveliotou, C., et al 1992a, IAU Circ. No. 5576

Kouveliotou, C., et al 1992b, IAU Circ. No. 5592

Kusunose, M., & Mineshige, S. 1994, ApJ, 423, 600

Lasota, J. P. & Pelat, D. 1991, A&A, 249, 574

Liang, E. P. T., & Price, R. H. 1977, ApJ, 218, 247

Lightman, A. P., & Eardley, D. N. 1974, ApJ, 187, L1

Lubin, L. M., Lewin, W. H. G., Rutledge, R. E., van Paradijs, J., van der Klis, M., & Stella, L. 1992, MNRAS, 258, 759

Lynden-Bell, D., & Pringle, J. E. 1974, MNRAS, 168, 603

Lyutyi, V. M. & Sunyaev, R. A. 1976, Sov. Ast., 20, 290

Matsumoto, R., Kato, S., & Honma, F. 1989, in Theory of Accretion Disks, ed. F. Meyer, W. J. Duschl, J. Frank, & E. Meyer-Hofmeister (Dordrecht: Kluwer), 167

Milsom, J. A., Chen, X., & Taam, R. E. 1994, ApJ, 421, 668

Nakamura, K., & Osaki, Y. 1993, PASJ, 45, 775

Narayan, R., & Popham, R. 1993, Nature, 362, 820

Narayan, R., & Yi, I. 1994, ApJ, 428, L13

Narayan, R., & Yi, I. 1995a, preprint

Narayan, R., & Yi, I. 1995b, preprint

Nowak, M. A., & Wagoner, R. V. 1992, ApJ, 393, 697

Pietsch, W., Haberl, F., Gehrels, N., & Petre, R. 1993, A&A, 273, L11

Ross, R. R., & Fabian, A. C. 1993, MNRAS, 261, 74

Shakura, N. I., & Sunyaev, R. A. 1973, A&A, 24, 337

Shapiro, S. L., Lightman, A. P., & Eardley, D. M. 1976, ApJ, 204, 187

Sunyaev, R. et al. 1991, A&A, , 247, L29

Svensson, R. & Zdziarski, A. 1994, ApJ, 436, 599

Taam, R. E., & Lin, D. N. C. 1984, ApJ, 287, 761

Taam, R. E., & Mészáros, P. 1987, ApJ, 322, 329

Tanaka, Y. 1991, in Iron Line Diagnostics in X-Ray Sources, Lecture Notes in Physics, ed. A. Treves, G. C. Perola, & L. Stella (New York: Springer-Verlag), 385, 98

Tanaka, Y. & Lewin, W. H. G., 1995, in X-Ray Binaries, ed. W. H. G. Lewin, J. van Paradijs & E. P. J. van den Heuvel (Cambridge: Cambridge University Press), in press

van der Klis, M. 1989, ARA&A, 27, 517

Vikhlinin, A., et al 1992a, IAU Circ. No. 5576

Vikhlinin, A., et al 1992b, IAU Circ. No. 5608

Vikhlinin, A., et al 1994, ApJ, 424, 395

Wallinder, F. H. 1991, A&A, 249, 107

– 13 –| Sequence | $M(M_\odot)$ | $\dot{M}(\dot{M}_c)$ | $\alpha$ | $r_{in}(r_g)$ | $\eta$ | $f$ (Hz) | $\Delta\dot{M}/\dot{M}_0$ |
|---|---|---|---|---|---|---|---|
| 1......... | 10.0 | 0.7 | 1.0 | 30 | 0.75 | 0.034 | 0.3 |
| 2......... | 10.0 | 0.65 | 1.0 | 30 | 0.75 | 0.038 | ... |
| 3......... | 10.0 | 0.75 | 1.0 | 30 | 0.75 | 0.032 | 0.5 |
| 4......... | 10.0 | 0.7 | 1.0 | 30 | 0.7 | 0.024 | 1.1 |
| 5......... | 10.0 | 0.7 | 1.0 | 20 | 0.75 | 0.029 | 1.0 |
| 6......... | 10.0 | 0.7 | 1.0 | 35 | 0.72 | 0.032 | 0.3 |
| 7......... | 1.4 | 0.3 | 0.02 | 8 | 0.5 | 0.064 | 0.5 |
| 8......... | 1.4 | 0.18 | 0.01 | 5 | 0.5 | 0.054 | 0.8 |
| 9......... | 1.4 | 0.3 | 0.01 | 7 | 0.5 | 0.039 | 0.55 |

Table 1: Model Sequences

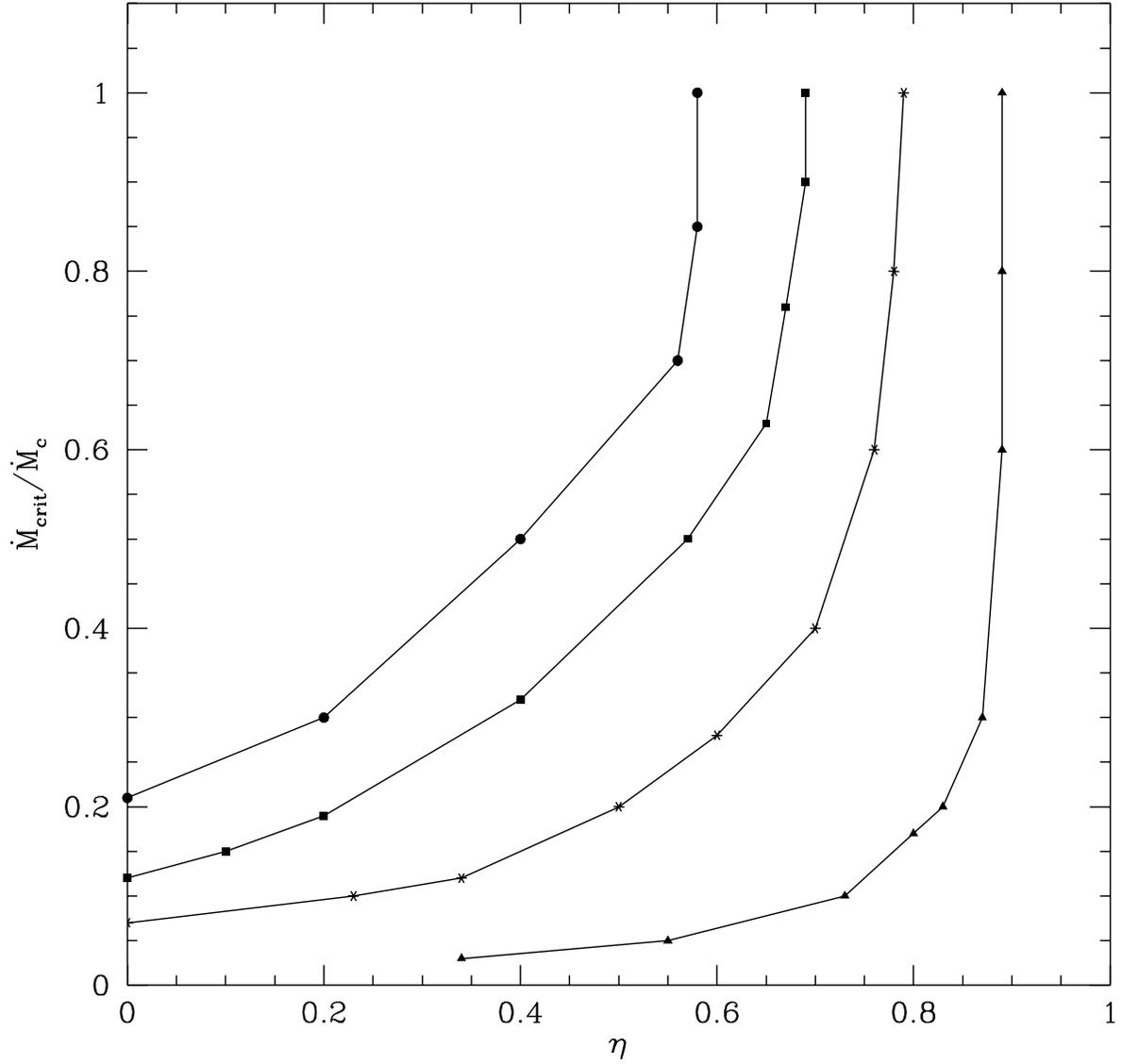

Fig. 1.— The $\dot{M}(\eta)$ relation where the cold disk is marginally stable for a specified inner radius with $M = 10 M_\odot$ and $\alpha = 1$. Here the solid circles, solid squares, solid asteriks, and solid triangles correspond to disks with inner radii of 70 $r_g$, 50 $r_g$, 30 $r_g$, and 10$r_g$ respectively. For parameters to the left of each curve the disk exhibits nonsteady mass flow modulations for a given inner disk radius. For parameters to the right of each curve the disk is globally stable for a given inner disk radius. The locations of model sequences 1, 2, 3, and 4 are illustrated for convenience.

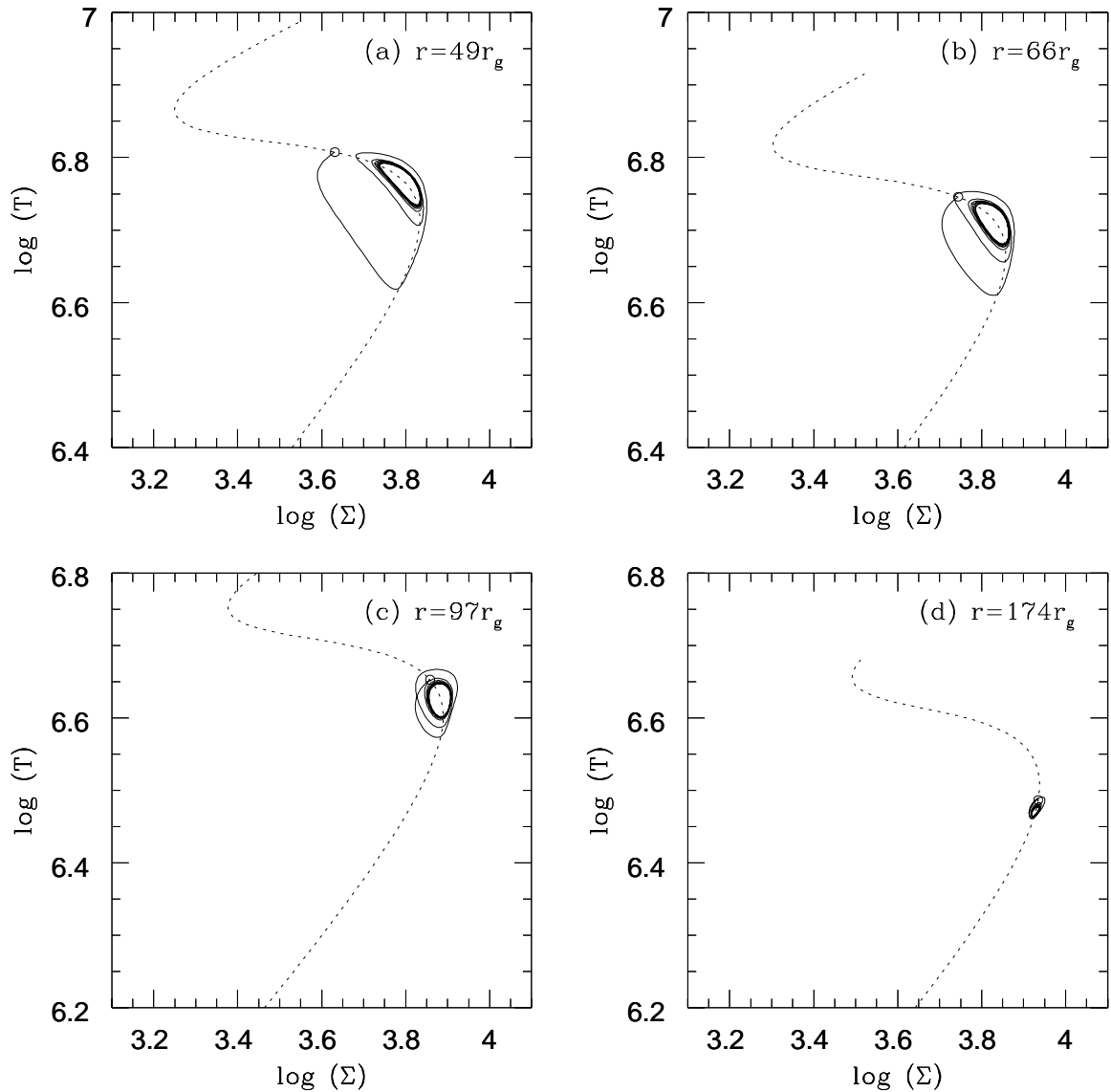

Fig. 2.— Evolutionary paths of the temperature and surface density of the standard sequence for four radii. The dashed curves denote the steady state S curve solutions and are shown for comparison. Note that the evolutionary paths are described by loops with larger excursions from the steady state curve for smaller radii.

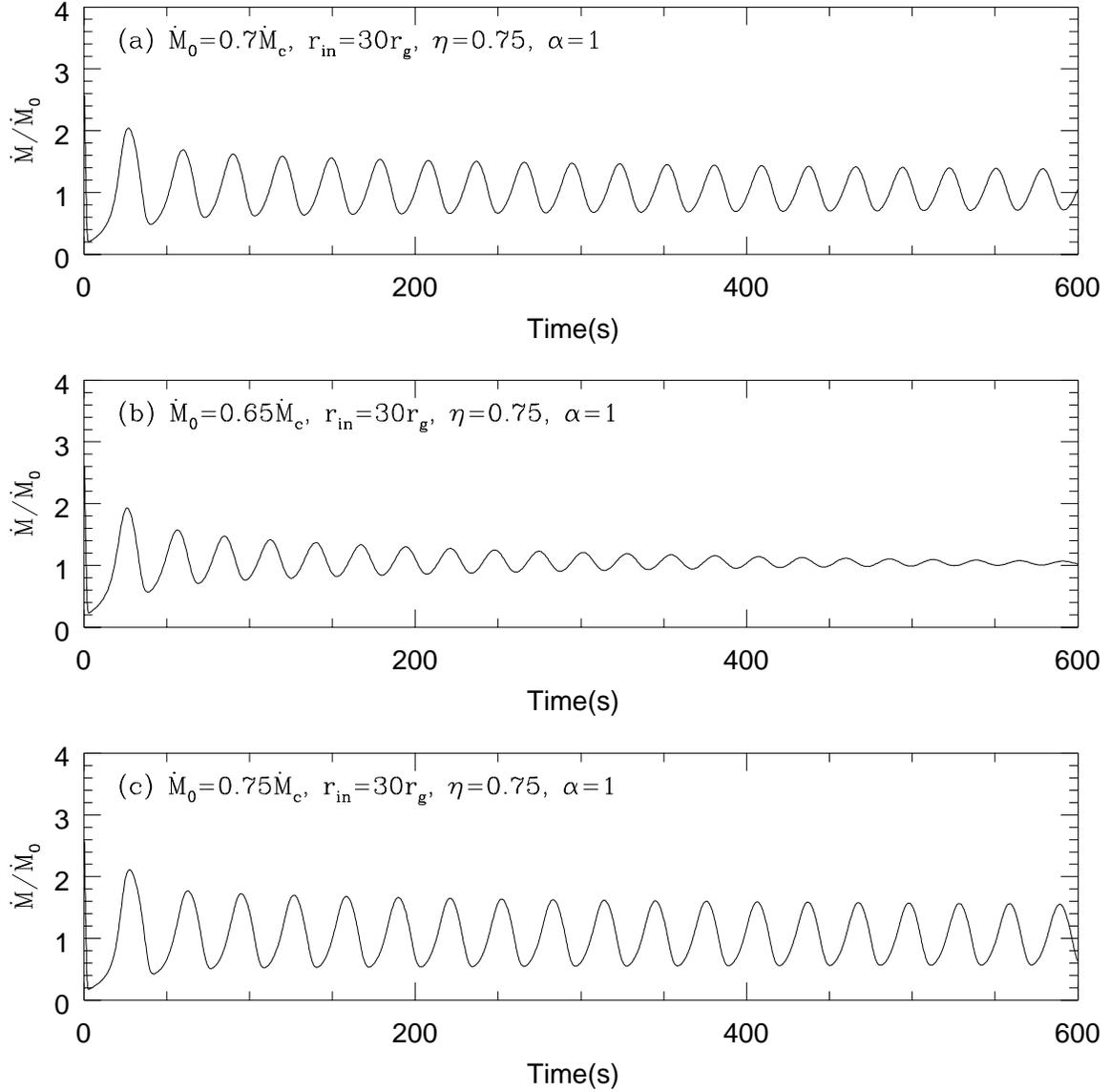

Fig. 3.— Time variations of the mass flow rate relative to the input rate, $\dot{M}_0$, at the inner edge of the disk. In all three sequences, $M = 10 M_\odot$, $\alpha = 1$, $\eta = 0.75$, and $r_{in} = 30 r_g$. The top (a), middle (b), and lower (c) panels correspond to sequences where $\dot{M} = 0.7$, $0.65$, and $0.75$ $\dot{M}_c$.

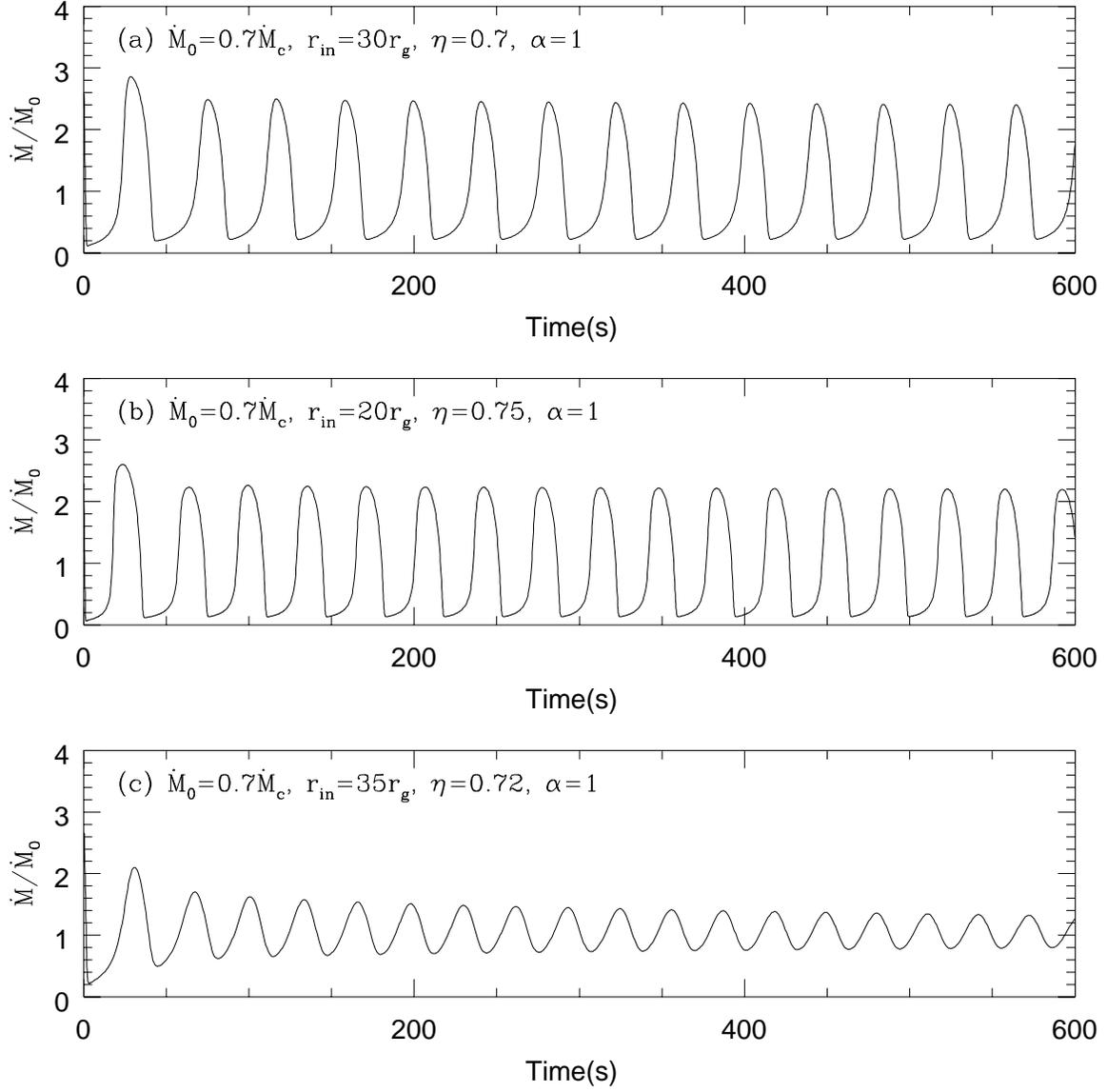

Fig. 4.— Variation of the disk mass flow rate at the inner edge of the disk relative to the input rate as a function of time. In sequences 4, 5, and 6 the central object is chosen to be a black hole with mass $M = 10 M_\odot$, the mass input rate is $\dot{M} = 0.7 \dot{M}_c$, and $\alpha = 1$. The top (a), middle (b), and lower (c) panels correspond to sequences 4, 5, and 6 respectively.

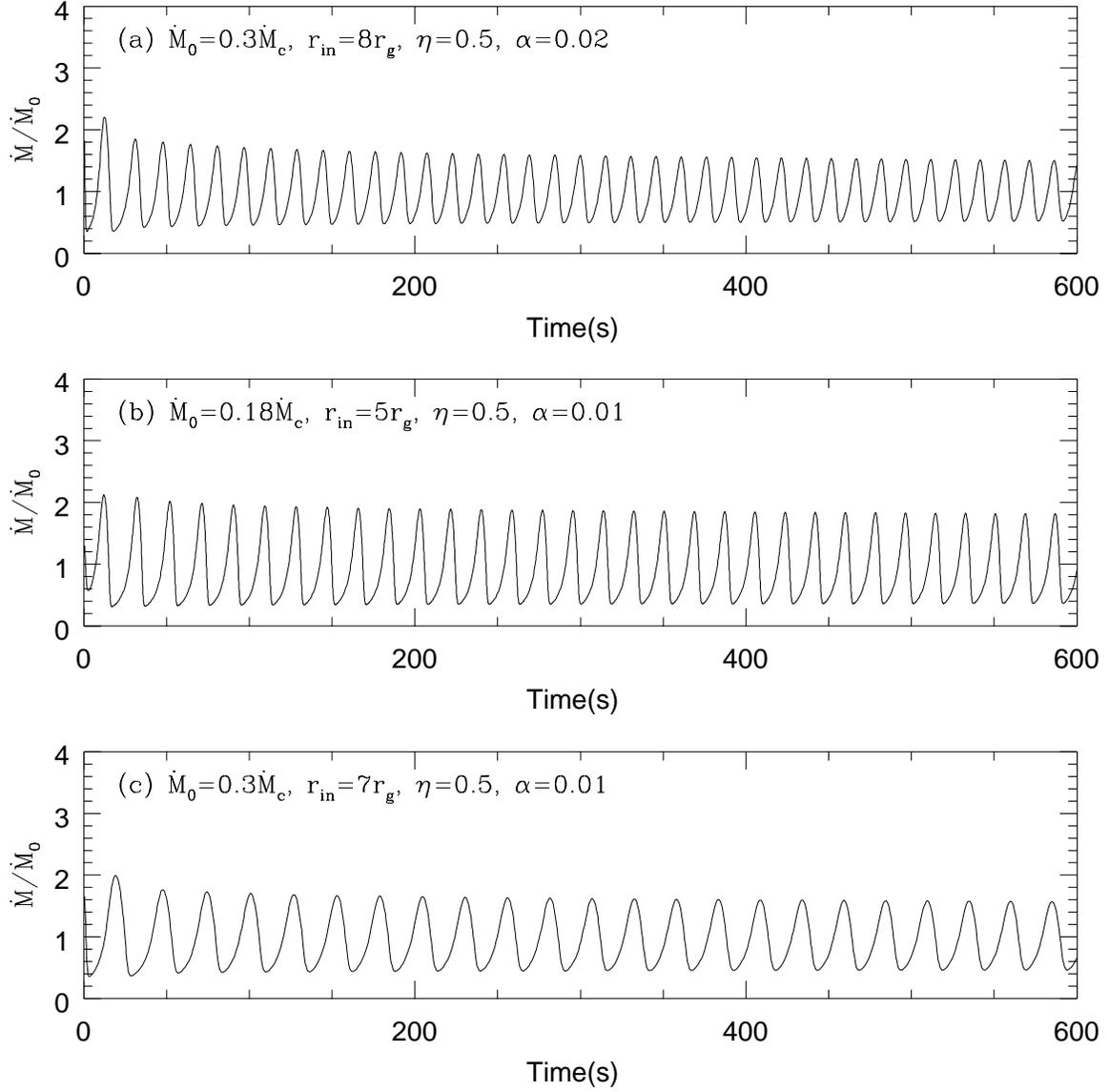

Fig. 5.— Temporal variability of the disk mass flow rate at the inner edge of the optically thick disk relative to the mass input rate. In all three sequences, the central object is chosen to be a neutron star with mass $M = 1.4 M_\odot$ and the fraction of energy dissipated in the corona is taken to be $\eta = 0.5$. The top (a), middle (b), and lower (c) panels correspond to sequences 7, 8, and 9 respectively.